\RequirePackage[2020-02-02]{latexrelease}
\documentclass[aps,prd,groupedaddress,graphicx,nofootinbib]{revtex4}
\usepackage{amsmath,amssymb,graphics,graphicx,color,epsf}
\usepackage{subfigure}
\usepackage[scanall]{psfrag}  
\usepackage{tikz}

\begin{document}

\title{Uniform Rate Inflation}

\author{Chia-Min Lin}

\affiliation{Fundamental General Education Center, National Chin-Yi University of Technology, Taichung 41170, Taiwan}



\begin{abstract}
In this work, I consider an inflation model with a quadratic potential and a negative cosmological constant. An analytical solution of the equation of motion for the inflaton field is found without slow-roll approximation. The result is that the inflation field is rolling at a constant speed. The prediction for cosmological perturbation is calculated.

\end{abstract}
\maketitle
\large
\baselineskip 18pt
\section{Introduction}

Cosmic inflation \cite{Starobinsky:1980te, Guth:1980zm, Linde:1981mu} is arguably the standard scenario for the very early universe cosmology. It refers to a period of accelerated expansion of the universe. This scenario solves many of the problems of the conventional hot big bang such as the flatness problem, the horizon problem, and the problem of unwanted relics. In addition, inflation provides superhorizon cosmological perturbations which can hardly be achieved without inflation. On the other hand, primordial density (curvature) perturbation and primordial gravitational waves observed or constrained through for example cosmic microwave background (CMB) or the large-scale structure turns out to be a powerful tool to test a model of cosmic inflation.

Inflation can be driven by the potential $V$ of one (or more) scalar field(s). The equation of motion of a homogeneous scalar field $\phi$ in an expanding universe is
\begin{equation}
\ddot{\phi}+3H\dot{\phi}+\partial_\phi V=0,
\label{momo}
\end{equation}
where $H$ is the Hubble parameter. One may solve this nonlinear equation numerically or use some approximation scheme such as slow-roll conditions. However, if it can be solved analytically, it is not only satisfying but also provides a unique chance to develop intuition, test the approximation scheme, and improve our understanding of the model or similar ones. One example in which the equation of motion can be solved analytically is power law inflation with the exponential potential \cite{Abbott:1984fp, Lucchin:1984yf, Lyth:1991bc}. 

In this work, we provide another example with a simple potential in which an analytical solution can be obtained. We consider a quadratic potential with a negative cosmological constant\footnote{One motivation to consider a negative cosmological constant is that Anti-de Sitter (AdS) vacua are ubiquitous in string theory. Sometimes it is even suggested that string theory has no de Sitter (dS) vacua \cite{Danielsson:2018ztv}.}.

\section{uniform rate inflation}

In the following discussion, we set the reduced Planck mass $M_P \simeq 2.4 \times 10^{18}\mbox{ GeV}$ to $M_P=1$.
Let us consider an inflaton field $\phi$ with the potential
\begin{equation}
V=V_0\left(\frac{3}{2}\phi^2-1 \right)=\frac{3}{4}\lambda^2\phi^2-\frac{1}{2}\lambda^2,
\label{po}
\end{equation}
where $V_0 \equiv \lambda^2/2$. 
We consider a flat universe and the scale factor $a$ is defined as
\begin{equation}
a \equiv e^\alpha.
\label{a}
\end{equation}
This implies the Hubble parameter is 
\begin{equation}
H \equiv \frac{\dot{a}}{a}=\dot{\alpha}.
\end{equation}
The Friedmann equation is given by
\begin{equation}
3\dot{\alpha}^2=3H^2=\frac{\dot{\phi}^2}{2}+V=\frac{\dot{\phi}^2}{2}+\frac{3}{4}\lambda^2 \phi^2 -\frac{\lambda^2}{2}.
\label{fri}
\end{equation}
The equation of motion of $\phi$ is
\begin{equation}
\ddot{\phi}+3H\dot{\phi}+\partial_\phi V=\ddot{\phi}+3\dot{\alpha}\dot{\phi}+\partial_\phi V=0.
\label{eom}
\end{equation} 
In the case of slow-roll inflation, one common way to deal with the equation is using the slow-roll approximation:
\begin{equation}
3H^2 \simeq V,  
\label{sr1}
\end{equation}
and 
\begin{equation}
3H\dot{\phi}+\partial_\phi V \simeq 0.
\label{sr2}
\end{equation}
This amounts to neglect of the term $\dot{\phi}^2/2$ in Eq.~(\ref{fri}) and $\ddot{\phi}$ in Eq.~(\ref{eom}). 
Interestingly, in our model, there is an analytical solution without appealing to slow-roll approximation. Since Eq.~(\ref{po}) is an even function of $\phi$, we can choose $\phi>0$ without loss of generality. The solution of Eq.~(\ref{eom}) is given by
\begin{equation}
\dot{\phi}=-\lambda,
\label{sol1}
\end{equation}
which can be integrated into $\phi=-\lambda t+c$ with an integration constant $c$. The Hubble parameter in Eq.~(\ref{fri}) is
\begin{equation}
H=\dot{\alpha}=\frac{\lambda \phi}{2}=-\frac{\lambda^2}{2}t+\frac{\lambda}{2}c,
\label{sol2}
\end{equation}
where we have used the integrated result of $\phi$ in the second equality. This can be integrated into $\alpha=-\frac{\lambda^2}{4}t^2+\frac{\lambda c}{2}t+d$ with an integration constant $d$.
We can freely set $d=0$ by imposing $\alpha(0)=0$ since only the ratios of the scale factors have physical meaning. 
The solution was originally found by the author when trying to solve the Wheeler-DeWitt equation through a study of quantum cosmology in the framework of de Broglie-Bohm theory in \cite{Lin:2023sza}. Usually, the evolution of the scale factor during slow-roll inflation is described as a ``quasi-exponential" expansion. Here we can see that the scale factor 
\begin{equation}
a=e^\alpha=e^{-\frac{\lambda^2}{4}t^2+\frac{\lambda c}{2}t}=e^{-\frac{\lambda^2}{4}\left( t-\frac{c}{\lambda} \right)^2+\frac{c^2}{4}}
\end{equation} 
is exactly given by a Gaussian function of time\footnote{However, as discussed in Appendix~\ref{b}, this Gaussian function behavior is unstable under perturbation during the contracting phase. The universe could evolve into a big crunch.}.
We call this model uniform rate inflation because the inflaton field is rolling at a constant speed\footnote{One might be tempted to call this model constant-roll inflation, but this terminology has been used to refer to the model of \cite{Motohashi:2014ppa} in a different context where $\ddot{\phi}$ is not negligible.}.

It can be easily checked by substituting Eqs.~(\ref{sol1}) and (\ref{sol2}) into Eqs.~(\ref{fri}) and (\ref{eom}) to see that they are exact solutions. Note that Eq.~(\ref{sr2}) becomes an equality (without approximation) in our model since $\ddot{\phi}=0$ in Eq.~(\ref{eom}). As we will see in the next section, the slow-roll condition Eq.~(\ref{sr1}) is also satisfied during inflation. As a bonus, this provides an opportunity to test the accuracy of the slow-roll approximation in this model. 

The particular solution given in Eq.~(\ref{sol1}) is just one out of many solutions. For example, the initial condition can be $\dot{\phi}=0$ which does not satisfy Eq.~(\ref{sol1}). What is special about it is that Eq.~(\ref{sol1}) is an attractor solution. This is not so unexpected since it is known that slow-rolling could be an attractor. We discuss this point more explicitly in the appendix.

\section{cosmological perturbations}

In order to test an inflation model, we should calculate the observables and compare them to experimental data.
Let us begin with an investigation of the scale factor $a$. From Eqs.~(\ref{a}) and (\ref{sol2}), we have
\begin{equation}
\dot{a}=e^\alpha \frac{\lambda \phi}{2}.
\end{equation}
This tells us that as long as $\phi>0$, the universe is expanding. The universe starts to collapse when $\phi=0$ is achieved. Taking one more derivative with respect to time, we obtain
\begin{equation}
\ddot{a}=e^\alpha \frac{\lambda}{2}\left( \frac{\lambda\phi^2}{2}+\dot{\phi} \right)=e^\alpha \frac{\lambda^2}{2}\left( \frac{\phi^2}{2}-1 \right),
\end{equation}
where we have used Eq.~(\ref{sol1}) in the second equality. This implies that when $\phi>\sqrt{2}$ (note that we use the unit $M_P=1$), we have $\ddot{a}>0$. This is the phase of inflation. The end of inflation happens when\footnote{If the slow-roll approximation is used to determine the end of inflation, it would be estimated to be $\phi=\sqrt{3.2}$.} $\phi = \phi_e =\sqrt{2}$. Our universe leaves the horizon at the number of e-folds $\Delta \alpha \simeq 60$. In order to calculate the corresponding field value $\phi=\phi_i$, we use Eqs.~(\ref{sol1}) and (\ref{sol2}) to obtain
\begin{equation}
\frac{d\alpha}{d\phi}=-\frac{\phi}{2}.
\label{para}
\end{equation}
Therefore 
\begin{equation}
\Delta \alpha = \int^{\phi_i}_{\sqrt{2}}\frac{\phi}{2}d\phi = \frac{\phi_i^2}{4}-\frac{1}{2} \simeq 60. 
\label{phi60}
\end{equation}
We may neglect $1/2$ compared with $60$ and obtain $\phi_i^2=240$. By using this result, one can easily check the slow-roll condition Eq.~(\ref{sr1}) is satisfied.

An ingenious way to calculate primordial density (curvature) perturbation is to use the $\delta N$ formalism \cite{Sasaki:1995aw, Sasaki:1998ug, Lyth:2004gb, Lyth:2005fi, Wands:2000dp}. Here we use $\alpha$ to denote $-N$, therefore
\begin{equation}
\delta N= -\delta \alpha =-\frac{d\alpha}{d\phi}\delta \phi=\frac{\phi}{2}\frac{H}{2\pi}=\frac{\lambda \phi^2}{8\pi},
\end{equation} 
where we have used Eqs.~(\ref{sol2}), (\ref{para}), and $\delta \phi=H/2\pi$.
The spectrum $P_R$ is given by
\begin{equation}
P_R=(\delta \alpha)^2=\frac{\lambda^2 \phi^4}{64\pi^2}.
\end{equation}
By imposing cosmic microwave background (CMB) normalization $P_R^{1/2}=5 \times 10^{-5}$ at $\phi_i^2=240$, we obtain
\begin{equation}
\lambda \simeq 5 \times 10^{-4}.
\end{equation}
The spectral index $n_s$ is
\begin{equation}
n_s \equiv 1+ \frac{d\ln P_R}{d\ln k}=1+ \frac{d\ln P_R}{d\alpha}=1-\frac{8}{\phi^2_i} \simeq 0.967,
\label{index}
\end{equation}
where $k$ is the comoving wave number of the perturbation. The running spectral index $n_s^\prime$ is
\begin{equation}
n_s^\prime \equiv \frac{dn_s}{d\ln k}=\frac{dn_s}{d\alpha}=\frac{32}{\phi_i^4} \simeq 5.56 \times 10^{-4}.
\end{equation}
The spectrum for the tensor perturbation is
\begin{equation}
P_T=8\left( \frac{H}{2\pi} \right)^2=\frac{2H^2}{\pi^2}.
\end{equation}
The tensor-to-scalar ratio is
\begin{equation}
r \equiv \frac{P_T}{P_R}=\frac{2}{\pi^2}\frac{\lambda^2 \phi^2_i}{4}\frac{64\pi^2}{\lambda^2 \phi^4_i}=\frac{32}{\phi_i^2} \simeq 0.13.
\label{r}
\end{equation}
From the observation of CMB by Planck \cite{Planck:2018jri}, the spectral index is given by $n_s=0.9649 \pm 0.0042$ at $68\%$ CL. The $95\%$ CL upper limit on the tensor-to-scalar ratio is $r<0.1$. The running spectral index is $n_s^\prime =-0.0045 \pm 0.0067$ at $68\%$ CL. As we can see, the fitting to the spectral index and its running is good. On the other hand, the tensor-to-scalar ratio is not favored by observation, but probably not definitely ruled out. We will see how to reduce $r$ in the next section. The model can be further tested in the future. The prediction of a large $r$ is expected since this model belongs to a large-field inflation model\footnote{Therefore, like all large-field models, it does not satisfy the swampland distance conjecture (see \cite{Agmon:2022thq} for a recent review and more references therein.).}. Actually, the overall predictions are similar to chaotic inflation \cite{Linde:1983gd} with a quadratic potential since when the field value is large the negative cosmological constant term in Eq.~(\ref{po}) can be neglected. But this does not mean solutions obtained from slow-roll approximation turn out to be exact.

\section{after inflation}
Usually, it is assumed that when the slow-roll conditions failed the inflaton field starts rolling rapidly and enters into an oscillating phase. However, in our model, the inflaton field is still rolling at a constant speed even after inflation!

\begin{figure}[t]
  \centering
\includegraphics[width=0.6\textwidth]{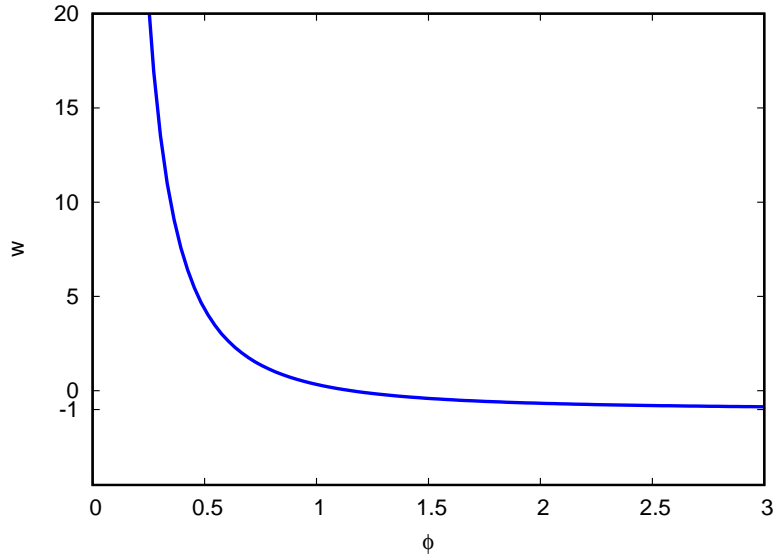}
  \caption{The equation of state $w$ as a function of the inflaton field value $\phi$ after inflation. At $\phi=\sqrt{2}$, we have $w=-\frac{1}{3}$. At $\phi=\frac{2}{\sqrt{3}}$, we have $w=0$. At large value of $\phi$, we have $w \rightarrow -1$.}
  \label{fig1}
\end{figure}

Let us investigate the equation of state for this peculiar behavior.
By using Eqs.~(\ref{po}) and (\ref{sol1}), the energy density of the inflaton field is given by
\begin{equation}
\rho=\frac{\dot{\phi}^2}{2}+V=\frac{3}{4}\lambda^2 \phi^2.
\end{equation}
Note that $\rho \geq 0$ although there is a negative cosmological constant. The pressure is given by
\begin{equation}
p=\frac{\dot{\phi}^2}{2}-V=\lambda^2-\frac{3}{4}\lambda^2 \phi^2.
\end{equation}
When $\phi=\frac{2}{\sqrt{3}}$, we have $p=0$ and the scalar field behaves like nonrelativistic matter. The equation of state $w$ is given by
\begin{equation}
w \equiv \frac{p}{\rho}=\frac{4-3\phi^2}{3\phi^2}.
\end{equation}
This is plotted in Fig.~\ref{fig1}. An interesting feature is when $\phi \rightarrow 0$, we have $w \rightarrow \infty$. For a realistic model, the inflaton may decay and reheat the universe at some time after inflation. Suppose the energy density of the inflaton field transforms to radiation. The total energy density of the universe is still positive. However, the radiation energy density is diluted during the expansion of the universe and the negative cosmological constant would slow down the expansion rate. Although the universe would never enter the phase of $\rho<0$ because before this happens, we would have $H \rightarrow 0$ and the universe stops its expansion and its dilution of radiation (or matter). This may happen when the universe is still quite hot! 

\begin{figure}[t]
  \centering
\includegraphics[width=0.6\textwidth]{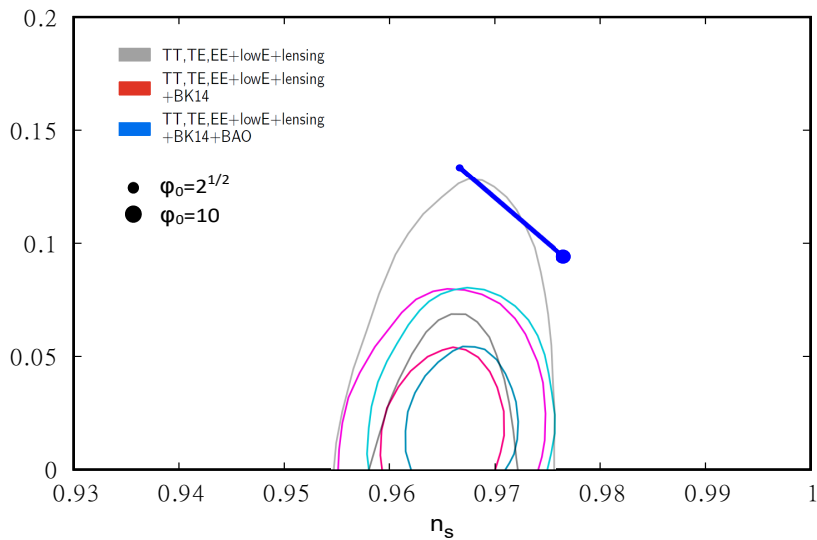}
  \caption{The spectral index $n_s$ and the tensor-to-scalar ratio $r$. The experimental constraints are from \cite{Planck:2018jri}.}
  \label{fig2}
\end{figure}

One solution is to assume the potential form in Eq.~(\ref{po}) only applies during inflation. The potential becomes negative at $\phi=\phi_c=\sqrt{2/3}$. At some point $\phi_0>\phi_c$, we may assume that the potential has been changed to another form without the negative cosmological constant. The inflaton field will then be able to roll rapidly, oscillate and decay to reheat the universe. Note that this is also necessary for power law inflation (mentioned in the introduction section), otherwise power law inflation cannot end. This idea introduces a new parameter $\phi_0$ into the model. Actually, we may even have $\phi_0>\sqrt{2}$ and assume that inflation ends at $\phi=\phi_0$. In this case, we will be able to reduce the tensor-to-scalar ratio $r$ at the expense of increasing $n_s$. For example, let us choose $\phi_0=10$. Instead of that given in Eq.~(\ref{phi60}), we would obtain $\phi_i^2=340$. Consequently, Eq.~(\ref{index}) becomes $n_s=0.976$ and Eq.~(\ref{r}) becomes $r=0.094$. The parameter $\lambda$ would reduce to $\lambda=3.7 \times 10^{-6}$. More generally, a relation
\begin{equation}
r=4(1-n_s)
\end{equation}
can be derived from Eqs.~(\ref{index}) and (\ref{r}). In order to compare the model with the latest experimental results, we plot the spectral index and the tensor-to-scalar ratio overlaid with the constraints taken from \cite{Planck:2018jri} in Fig.~\ref{fig2}.

\section{conclusion}
\label{con}
In this work, we propose an inflation model with a simple potential that admits an analytical solution during inflation. Although inflation can end automatically in our model, it will start to collapse soon after inflation due to the negative cosmological constant. We introduce a new parameter $\phi_0$ to signify that the potential form given in Eq.~(\ref{po}) is only valid when $\phi>\phi_0$. 
If $\phi_0 \leq \sqrt{2}$, inflation ends at $\phi_e=\sqrt{2}$ and we would have $\lambda \simeq 5 \times 10^{-4}$ via CMB normalization. Subsequently, it predicts $n_s=0.967$, $n_s^\prime \simeq 5.56 \times 10^{-4}$, and $r=0.13$. The tensor-to-scalar ratio $r$ can be further reduced at the expense of increasing $n_s$ if inflation ends somewhere at $\phi_0>\sqrt{2}$.
These results are obtained without slow-roll approximation. The behavior of the scale factor when $\phi>\phi_0$ is a Gaussian function. 


\appendix
\section{stability of the solution}
In this section, we consider the question about what would happen if $\dot{\phi}$ deviates from that given in Eq.~(\ref{sol1}).
Intuitively, we can see from Eqs.~(\ref{fri}) and (\ref{eom}) that if somehow $\dot{\phi}$ is increased (decreased) (with the same $V$ and $\partial_\phi V$), $H$ would also increase (decrease). Therefore the term $3H \dot{\phi}$ in Eq.~(\ref{eom}) is increased (decreased). In order to compensate for this increasing (decreasing), a negative (positive) $\ddot{\phi}$ appears and this reduces (enhances) $\dot{\phi}$ towards the value given by Eq.~(\ref{sol1}). A feedback mechanism is working here.

We can be more precise.
Let us consider a perturbation $u$ of the solution 
\begin{eqnarray}
\dot{\phi} &\rightarrow& \dot{\phi}+u,  \\
\ddot{\phi}&\rightarrow& \ddot{\phi}+\dot{u}\\
\dot{\alpha}&\rightarrow& \dot{\alpha}+\delta \dot{\alpha}.
\end{eqnarray}
Substituting into Eq.~(\ref{fri}) to obtain
\begin{equation}
3(\dot{\alpha}+\delta \dot{\alpha})^2=\frac{(\dot{\phi}+u)^2}{2}+V.
\end{equation} 
Deleting second-order terms, and using the zeroth order equation. We obtain
\begin{equation}
\delta \dot{\alpha}=-\frac{u}{3\phi}.
\end{equation}
From Eq.~(\ref{eom}) and the above equation, we have
\begin{equation}
\ddot{\phi}+\dot{u}+3(\dot{\alpha}+\delta\dot{\alpha})(\dot{\phi}+u)+\partial_\phi V=0.
\end{equation}
By using $\phi=c-\lambda t$, Eq.~(\ref{sol2}), and taking first-order perturbation, we have
\begin{equation}
\dot{u}+\frac{3\lambda \phi}{2}u+\frac{\lambda}{\phi}u=\dot{u}+\frac{3\lambda}{2}(c-\lambda t)u+\frac{\lambda}{c-\lambda t}u=0.
\end{equation}
The above equation can be integrated to obtain
\begin{equation}
u=be^{\ln|c-\lambda t|+\frac{3}{4}(c-\lambda t)^2}=b|\phi|e^{\frac{3}{4}\phi^2},
\end{equation}
where $b$ is an integration constant. Note that $u$ is a small perturbation and $\phi$ is decreasing during inflation. Therefore $u$ drops out very fast. For example, if $\phi^2$ goes from $280$ to $240$. The perturbation $u$ drops by a factor of $\sim e^{-30}$. 

\section{the big crunch}
\label{b}

Due to the negative cosmological constant, there must also exist solutions that result in a big crunch. 
When $\phi<0$, the universe is in a contracting phase with a negative Hubble parameter. The scale factor evolves like a time reversal of the expansion phase. Therefore the attractor becomes a repeller.
Let us consider the equation of motion of Eq.~(\ref{momo}). Now we have $\dot{\phi}<0$ and $H<0$.
If there is a small perturbation $\delta \dot{\phi}<0$ (similar consideration can be made for $\delta \dot{\phi}>0$), it makes $3H \dot{\phi}$ increase and this makes $\ddot{\phi}<0$ which increases $|\dot{\phi}|$ further and so on. If this keeps on going, we would have $\dot{\phi}^2/2 \gg V$ and $|\ddot{\phi}| \gg |V^\prime|$ which means the universe enters into a kination phase. Let us assume $x \equiv \dot{\phi}$ during kination. We have $H=x/\sqrt{6}$ and the equation of motion becomes
\begin{equation}
\frac{dx}{dt}=-\frac{3}{\sqrt{6}}x^2.
\end{equation}
This can be easily integrated to give
\begin{equation}
x=\frac{\sqrt{6}}{3\left( t-t_0 \right)},
\end{equation}
where $t_0$ is an integration constant. We have $t<t_0$ because $x<0$. Now the Hubble parameter is
\begin{equation}
H=\frac{\dot{a}}{a}=\frac{1}{3\left( t-t_0 \right)}.
\end{equation}
This can be integrated to give
\begin{equation}
a \propto \left( t_0-t \right)^{\frac{1}{3}}.
\end{equation}
When $t \rightarrow t_0$, we have $a \rightarrow 0$. This is the big crunch.


\acknowledgments
This work is supported by the National Science and Technology Council (NSTC) of Taiwan under Grant No. NSTC 111-2112-M-167-002.

\end{document}